\documentclass[preprintnumbers,amsmath,amssymb,prd,floatfix,12pt,superscriptaddress,nofootinbib]{revtex4}
\usepackage{graphicx}
\usepackage{epsfig}
\usepackage{bm}
\usepackage{amsfonts}
\usepackage[colorlinks]{hyperref}
\usepackage{float}

\newcommand{\be}{\begin{equation}}
\newcommand{\ee}{\end{equation}}
\newcommand{\ba}{\begin{eqnarray}}
\newcommand{\ea}{\end{eqnarray}}
\newcommand{\bal}{\begin{align}}
\newcommand{\eal}{\end{align}}

\newcommand{\bw}{\begin{widetext}}
\newcommand{\ew}{\end{widetext}}

\begin{document}

\title{Noether symmetries and first integrals of damped harmonic oscillator}

\author{M. Umar Farooq}
\email{m\_ufarooq@yahoo.com}
\affiliation{Department of Basic Sciences $\&$ Humanities, College of E $\&$ ME, National University of Sciences and Technology (NUST), H-12, Islamabad Pakistan}
\author{M. Safdar}
\email{safdar.camp@gmail.com}
\affiliation{School of Mechanical $\&$ Manufacturing Engineering, National University of Sciences and Technology (NUST), H-12, Islamabad Pakistan}

\begin{abstract}
	\vspace*{1.5cm} \centerline{\bf Abstract} \vspace*{1cm} Noether's theorem establishes an interesting connection between symmetries of the action integral and conservation laws (first integrals) of a dynamical system. The aim of the present work is to classify the damped harmonic oscillator problem with respect to Noether symmetries and to construct corresponding conservation laws for all over-damped, under-damped and critical damped cases. For each case we obtain maximum five linearly independent group generators which provide related five conserved quantities. Remarkably, after obtaining complete set of invariant quantities we obtain analytical solutions for each case. In the current work, we also introduce a new Lagrangian for the damped harmonic oscillator. Though the form of this new Lagrangian and presented by Bateman are completely different, yet it generates same set of Noether symmetries and conserved quantities. So, this new form of Lagrangian we are presenting here may be seriously interesting for the physicists. Moreover, we also find the Lie algebras of Noether symmetries and point out some interesting aspects of results related to Noether symmetries and first integrals of damped harmonic oscillator which perhaps not reported in the earlier studies.\\
	Key words: Damped harmonic oscillator, Noether symmetries, conservation laws.
\end{abstract}


\maketitle

\newpage
\section{Introduction}
The study of Noether symmetries and corresponding first integrals of underlying dynamical system has important mathematical significance as well as it has profound physical background. Discussion of Lie point transformations which leave the action integral of the damped harmonic oscillator (DHO) invariant has widened in recent years. Among the enormous and useful results concerning DHO, the discussion of Noether symmetries and the relevant conserved quantities have attracted attention of many researchers. If a Lie point transformation leaves the action integral invariant, the classical Noether's theorem \cite{Noe} provides a corresponding first integral. Another significance of these conserved quantities is that they help us in establishing solutions of underlying systems. Moreover, Lagrangian and Hamiltonian functions play essential role in deriving equations of motion. Therefore, in order to study relation of symmetries and conservation laws using Noether's theorem, one needs the corresponding equation of motion that arise from the action principle \cite{Gold}. A number of studies have been reported to construct integral of motions of DHO \cite{Cho,And,Cer}. In the current study, we intend to investigate the symmetry properties of DHO equation that is derivable from a variational principle. We shall see that the full eight parameter group that leaves the equation of motion invariant, only five of them leave the action integral invariant and become Noether symmetries. Consequently, we obtain only five conserved quantities for DHO. It is commonly known that maximum number of conserved quantities of a dynamical system often leads to find the solution of underlying equation. In this paper, we find many results on Noether symmetries and conservation laws of DHO and point out some interesting results related to Noether symmetries and first integrals to that already available in the literature. We also present and discuss a new form of Lagragian (may be taken as standard Lagragian) for DHO system which has not been reported so far. Though the new form of Lagrangian is completely different yet it yields same results as derived from the Bateman Lagrangian. So this new form of optimal Lagragian function may conceal something interesting for the physicists.\\
A brief outline of the present work is as follows: In Section II, we outline our treatment on Noether symmetries and Noether's theorem. In Section III, we give a classification of DHO with respect to Noether symmetries and construct correspondent conserved quantities for three cases. We provide Lie algebras of Noether operators and also discuss new form of Lagrangian for DHO. Finally, we present a brief summary of our work in Section IV.
\section{Preliminaries on Noether Symmetries and First integrals}
In order to present results of the current study in a straightforward manner, we start with reviewing expressions on finding Noether symmetry generators and associated conservation laws for DHO. We shall use the salient features of this section to derive important results in the next section. \\
Consider the point type vector field
\begin{eqnarray}
Z=\xi(t,u)\partial_{t}+\eta(t,u)\partial_{u},
\end{eqnarray}
then its first prolongation can be expressed as
\begin{equation}
Z^{[1]}=Z+(\eta_{t}+\eta_{u}u^{\prime}-\xi_{t}u^{\prime}-\xi_{u}u^{\prime
	2})\partial_{u^{\prime}}.
\end{equation}
Further assume that we have a first-order Lagrangian function $L(t,u,u')$ whose insertion in the Euler-Lagrange equation
\begin{eqnarray}
\frac{d}{dt}(\frac{\partial L}{\partial u^{\prime}})-\frac{\partial
	L}{\partial u}=0,
\end{eqnarray}
yields the following equation of motion
\begin{eqnarray}
u^{\prime\prime}=w(t,u,u^{\prime}).
\end{eqnarray}
We recall that the vector field $Z$ is called a Noether point symmetry generator corresponding to the Lagrangian $L(t,u,u')$ if a gauge function $B(t,u)$ exists such that 
\begin{eqnarray}
Z^{[1]}(L)+D(\xi)L=D(B),
\end{eqnarray}
holds, where
\begin{eqnarray}
D=\partial_{t}+u^{\prime}\partial_{u}+u^{\prime\prime}\partial_{u^{\prime}}+...  .
\end{eqnarray}
The significance of a Noether symmetry is that it provides the first integral as well as double reduction of order of underlying equation of motion.\\
\textbf{Noether's Theorem:} Suppose that a Lie point symmetry generator $Z$ satisfies the the condition given in Eq. (5), then the following expression leads to the conserved quantity or constant of motion of (4) related to $Z$.
\begin{eqnarray}
I=\xi L+(\eta-u^{\prime}\xi)\frac{\partial L}{\partial
	u^{\prime}}-B.
\end{eqnarray}
\section{Noether symmetries and first integrals of damped harmonic oscillator}
In this section we wish to find Noether symmetries and then utilizing them in the Noether's theorem to determine first integrals of a famous dynamical equation depicting motion of DHO. We shall classify the problem with respect to Noether's symmetries and construct the associated conserved quantities for over-damped, under-damped and critically damped cases and also determine their Lie algebras. So for this we proceed as:\\
Consider the physical system of DHO whose dynamics is governed by the following second-order ordinary differential equation
\begin{eqnarray}
u^{\prime\prime}+\frac{c}{m}u^{\prime}+\frac{k}{m}u=0.\label{eq1}
\end{eqnarray}
The well known time dependent Lagrangian of above Eq. (8) is given by the so called Bateman Lagrangian \cite{Bat,Yan,Edw,Ray}
\begin{eqnarray}
L=\frac{1}{2}\exp\left({\frac{ct}{m}}\right)(mu^{\prime 2}-ku^{2}),\label{lag1}.
\end{eqnarray}
From Eq. (9), it is evident that the time dependency of Lagrangian does not reveal the energy conservation of the system as one observes in the usual harmonic oscillator case. After inserting the above Lagrangian in eq. (5) and solving the resulting set of partial differential equations, we obtain the following Noether symmetries
\begin{equation}\label{p13}
\begin{aligned}
\textbf{X}_{1}&=\frac{\partial}{\partial t}-\frac{cu}{2m}\frac{\partial}{\partial u},\\
\textbf{X}_{2}&=\sin\Big(\frac{\sqrt{4km-c^2}t}{m}\Big)\frac{\partial}{\partial t}
+\Big(\frac{u\sqrt{4km-c^2}}{2m}\cos\Big(\frac{\sqrt{4km-c^2}t}{m}\Big)-\frac{cu}{2m}\sin\Big(\frac{\sqrt{4km-c^2}t}{m}\Big)\Big)\frac{\partial}{\partial u},\\
\textbf{X}_{3}&=\cos\Big(\frac{\sqrt{4km-c^2}t}{m}\Big)\frac{\partial}{\partial t}-\Big(\frac{u\sqrt{4km-c^2}}{2m}\sin\Big(\frac{\sqrt{4km-c^2}t}{m}\Big)+\frac{cu}{2m}\cos\Big(\frac{\sqrt{4km-c^2}t}{m}\Big)\Big)\frac{\partial}{\partial u},\\
\textbf{X}_{4}&=\exp\left(\frac{(-c-\sqrt{c^2-4km})t}{2m}\right)\frac{\partial}{\partial u},\\
\textbf{X}_{5}&=\exp\Big(\frac{(-c+\sqrt{c^2-4km})t}{2m}\Big)\frac{\partial}{\partial u}.
\end{aligned}
\end{equation}
Now by employing these operators in the Noether's theorem gives rise respectively the following 5-first integrals
\begin{equation}
\begin{aligned}
I_{1}&=-\exp\Big({\frac{ct}{m}}\Big)(mu^{\prime 2}+cuu^{\prime}+ku^{2}),\\
I_{2}&=-\frac{\exp\Big({\frac{ct}{m}}\Big)}{4m}(2m^2u^{\prime 2}+2mu(cu^{\prime}-ku)+c^2u^2)\sin\Big(\frac{\sqrt{4km-c^2}t}{m}\Big)\\ &+\frac{\exp\Big({\frac{ct}{m}}\Big)}{4m}u\sqrt{4km-c^2}(cu+2mu^{\prime})\cos\Big(\frac{\sqrt{4km-c^2}t}{m}\Big),\\
I_{3}&=-\frac{\exp\Big({\frac{ct}{m}}\Big)}{4m}(2m^2u^{\prime 2}+2mu(cu^{\prime}-ku)+c^2u^2)\cos\Big(\frac{\sqrt{4km-c^2}t}{m}\Big)\\
&-\frac{\exp\Big({\frac{ct}{m}}\Big)}{4m}u\sqrt{4km-c^2}(cu+2mu^{\prime})\sin\Big(\frac{\sqrt{4km-c^2}t}{m}\Big),\\
I_{4}&=\exp\Big({\frac{ct}{m}}\Big)\exp\Big({\frac{-(c+\sqrt{c^2-4km})t}{2m}}\Big)mu^{\prime}+\frac{1}{2}(c+\sqrt{c^2-4km})u\exp\Big({\frac{(c-\sqrt{c^2-4km})t}{2m}}\Big),\\
I_{5}&=\exp\Big({\frac{ct}{m}}\Big)\exp\Big({\frac{(-c+\sqrt{c^2-4km})t}{2m}}\Big)mu^{\prime}+\frac{1}{2}(c-\sqrt{c^2-4km})u\exp\Big({\frac{(c+\sqrt{c^2-4km})t}{2m}}\Big).
\end{aligned}
\end{equation}
Henceforth, we shall discuss 3-forms of DHO which include over-damped, under-damped and critical damped. We shall use conditions for each form and retrieve some known as well as some new results.\\
\textbf{Case 1. \textbf{Over-damped oscillator:}}\\ If the damping force is much stronger than the restoring force of the system, then it is called the over-damped harmonic oscillator. In this case $4km<c^2$, and the set of generators given in Eq. (10) takes the form
\begin{equation}\label{genset2}
\begin{aligned}
\textbf{X}_{1}&=\frac{\partial}{\partial t}-\frac{cu}{2m}\frac{\partial}{\partial u},\\
\textbf{X}_{2}&=\iota\sinh\Big(\frac{\sqrt{c^2-4km}t}{m}\Big)\frac{\partial}{\partial t}+\iota\Big(\frac{u\sqrt{c^2-4km}}{2m}\cosh\Big(\frac{\sqrt{c^2-4km}t}{m}\Big)-\frac{cu}{2m}\sinh\Big(\frac{\sqrt{c^2-4km}t}{m}\Big)\Big)\frac{\partial}{\partial u},\\
\textbf{X}_{3}&=\cosh\Big(\frac{\sqrt{c^2-4km}t}{m}\Big)\frac{\partial}{\partial t}-\Big(\frac{u\sqrt{c^2-4km}}{2m}\sinh\Big(\frac{\sqrt{c^2-4km}t}{m}\Big)+\frac{cu}{2m}\cosh\Big(\frac{\sqrt{c^2-4km}t}{m}\Big)\Big)\frac{\partial}{\partial u},\\
\textbf{X}_{4}&=\exp\Big(\frac{(-c-\sqrt{c^2-4km})t}{2m}\Big)\frac{\partial}{\partial u},\\
\textbf{X}_{5}&=\exp\Big(\frac{(-c+\sqrt{c^2-4km})t}{2m}\Big)\frac{\partial}{\partial u}.
\end{aligned}
\end{equation}
On closely viewing these operators, we note that the Noether operators  $\textbf{X}_{2}$ and $\textbf{X}_{3}$ attain different forms to those given in Eq. (10) while the other three generators $\textbf{X}_{1}$, $\textbf{X}_{4}$, $\textbf{X}_{5}$ and the associated conserved quantities ${I}_{1}$, ${I}_{4}$, ${I}_{5}$ remained unchanged. Hence, from the constants of the motion ${I}_{4}$ and ${I}_{5}$ we obtain the following solution for the over-damped oscillator
\begin{equation}
u(x)=\frac{2m}{c^2-4km}\Big[I_4e^\frac{-c+\sqrt{c^2-4km}}{2m}-I_5e^\frac{-c-\sqrt{c^2-4km}}{2m}\Big].
\end{equation}
Moreover, by utilizing the Noether symmetries $\textbf{X}_{2}$ and $\textbf{X}_{3}$ along with the Lagrangian (9) in Noether's theorem, we  determine respectively the following two new correspondent conserved quantities
\begin{equation}
\begin{aligned}
I_{2}&=\frac{1}{16km^2}\Big(-4km\exp\Big(\frac{ct}{m}\Big)(2m^2u^{\prime 2}+2m(cuu^{\prime}-ku^{2})+c^2u^2)\sinh\Big(\frac{\sqrt{c^2-4km}t}{m}\Big)+\\&\exp\Big(\frac{ct}{m}\Big)(cu^2+2muu^{\prime})(c^2\sqrt{c^2-4km}-(c^2-4km)^{3/2})\cosh\Big(\frac{\sqrt{c^2-4km}t}{m}\Big)\Big),\\
I_{3}&=\frac{1}{16km^2}\Big(-4km\exp\Big(\frac{ct}{m}\Big)(2m^2u^{\prime 2}+2m(cuu^{\prime}-ku^{2})+c^2u^2)\cosh\Big(\frac{\sqrt{c^2-4km}t}{m}\Big)+\\&\exp\Big(\frac{ct}{m}\Big)(cu^2+2muu^{\prime})(c^2\sqrt{c^2-4km}-(c^2-4km)^{3/2})\sinh\Big(\frac{\sqrt{c^2-4km}t}{m}\Big)\Big).
\end{aligned}
\end{equation}
Furthermore, the Lie algebra of these operators is shown in the following table
\begin{table}[h!]
	\centering
	\begin{tabular}{c| c c c c c} 
		&$\textbf{X}_{1}$&$\textbf{X}_{2}$ & $\textbf{X}_{3}$ & $\textbf{X}_{4}$ & $\textbf{X}_{5}$ \\ 
		\hline
		\hline
		$\textbf{X}_{1}$ & $0$ & $\frac{\sqrt{c^2-4km}}{m}\textbf{X}_{3}$ & $\frac{\sqrt{c^2-4km}}{m}\textbf{X}_{2}$ & $-\frac{\sqrt{c^2-4km}}{2m}\textbf{X}_{4}$ &$\frac{\sqrt{c^2-4km}}{2m}\textbf{X}_{5}$ \\
		
		$\textbf{X}_{2}$ &$-\frac{\sqrt{c^2-4km}}{m}\textbf{X}_{3}$&$0$& $-\frac{\sqrt{c^2-4km}}{m}\textbf{X}_{1}$ & $-\frac{\sqrt{c^2-4km}}{2m}\textbf{X}_{5}$ &$-\frac{\sqrt{c^2-4km}}{2m}\textbf{X}_{4}$ \\
		
		$\textbf{X}_{3}$ &$-\frac{\sqrt{c^2-4km}}{m}\textbf{X}_{2}$ & $\frac{\sqrt{c^2-4km}}{m}\textbf{X}_{1}$ & $0$ &$-\frac{\sqrt{c^2-4km}}{2m}\textbf{X}_{5}$ & $\frac{\sqrt{c^2-4km}}{2m}\textbf{X}_{4}$\\
		
		$\textbf{X}_{4}$ & $\frac{\sqrt{c^2-4km}}{2m}\textbf{X}_{4}$ & $\frac{\sqrt{c^2-4km}}{2m}\textbf{X}_{5}$ & $\frac{\sqrt{c^2-4km}}{2m}\textbf{X}_{5}$ &$0$ &$0$ \\
		
		$\textbf{X}_{5}$ & $-\frac{\sqrt{c^2-4km}}{2m}\textbf{X}_{5}$ & $\frac{\sqrt{c^2-4km}}{2m}\textbf{X}_{4}$ & $-\frac{\sqrt{c^2-4km}}{2m}\textbf{X}_{4}$ &$0$ & $0$\\ [1ex]
	\end{tabular}
	\caption{Commutators Table}
	\label{table:1}
\end{table}\\
The list of generators provided in Eq. (12) except $X_1$ is entirely different from that presented in \cite{Cer}. Consequently, the associated first integrals are also non-identical to those mentioned in \cite{Cer}. Moreover, from above we see that if we put $c=0$, the operator $X_1$ becomes time translational symmetry which yields integral $I_1$ as energy conservation of the usual harmonic oscillator. Furthermore, under the assumption that the system is free of damping term, all the above symmetry generators and corresponding first integrals become identical to what determine by Lutzky \cite{Lut}.\\ 
\textbf{Case 2. Under-damped Oscillator:}\\ If the restoring force is large compared to the damping force term, then the system is termed as the under-damped. This situation occurs when $4km>c^2$, so the resulting Noether symmetry generators are expressed as
\begin{equation}\label{genset3}
\begin{aligned}
\textbf{X}_{1}&=\frac{\partial}{\partial t}-\frac{cu}{2m}\frac{\partial}{\partial u},\\
\textbf{X}_{2}&=\sin\Big(\frac{\sqrt{4km-c^2}t}{m}\Big)\frac{\partial}{\partial t}+\Big(\frac{u\sqrt{4km-c^2}}{2m}\cos\Big(\frac{\sqrt{4km-c^2}t}{m}\Big)-\frac{cu}{2m}\sin\Big(\frac{\sqrt{4km-c^2}t}{m}\Big)\Big)\frac{\partial}{\partial u},\\
\textbf{X}_{3}&=\cos\Big(\frac{\sqrt{4km-c^2}t}{m}\Big)\frac{\partial}{\partial t}-\Big(\frac{u\sqrt{4km-c^2}}{2m}\sin\Big(\frac{\sqrt{4km-c^2}t}{m}\Big)+\frac{cu}{2m}\cos\Big(\frac{\sqrt{4km-c^2}t}{m}\Big)\Big)\frac{\partial}{\partial u},\\
\textbf{X}_{4}&=\exp\Big(\frac{-ct}{2m}\Big)\Big(\cos\Big(\frac{\sqrt{4km-c^2}t}{2m}\Big)-\iota \sin\Big(\frac{\sqrt{4km-c^2}t}{2m}\Big)\Big)\frac{\partial}{\partial u},\\
\textbf{X}_{5}&=\exp\Big(\frac{-ct}{2m}\Big)\Big(\cos\Big(\frac{\sqrt{4km-c^2}t}{2m}\Big)+\iota \sin\Big(\frac{\sqrt{4km-c^2}t}{2m}\Big)\Big)\frac{\partial}{\partial u}.
\end{aligned}
\end{equation}
In this case the operators $\textbf{X}_{4}$ and $\textbf{X}_{5}$, are different from that shown in (10). Interestingly, the splitting one of these two symmetries $\textbf{X}_{4}$ and $\textbf{X}_{5}$ into their real and imaginary parts results in two new Noether symmetries as shown
\begin{equation}
\begin{aligned}
\textbf{G}_{4}&=\exp\Big(\frac{-ct}{2m}\Big)\cos\Big(\frac{\sqrt{4km-c^2}t}{2m}\Big)\frac{\partial}{\partial u},\\
\textbf{G}_{5}&=\exp\Big(\frac{-ct}{2m}\Big)\sin\Big(\frac{\sqrt{4km-c^2}t}{2m}\Big)\frac{\partial}{\partial u}.
\end{aligned}
\end{equation}
Now insertion of the Lagrangian (9) along with Noether symmetries mentioned in Eq. (16), in Noether's theorem (7) leads towards following pair of conserved quantities
\begin{equation}
\begin{aligned}
I_{4}&=\exp\Big(\frac{ct}{2m}\Big)\Big[(mu^{\prime}+\frac{cu}{2})\cos\Big(\frac{\sqrt{4km-c^2}t}{2m}\Big)+\frac{1}{2}u\sqrt{4km-c^2}\sin\Big(\frac{\sqrt{4km-c^2}t}{2m}\Big)\Big],\\
I_{5}&=\exp\Big(\frac{ct}{2m}\Big)\Big[(mu^{\prime}+\frac{cu}{2})\sin\Big(\frac{\sqrt{4km-c^2}t}{2m}\Big)-\frac{1}{2}u\sqrt{4km-c^2}\cos\Big(\frac{\sqrt{4km-c^2}t}{2m}\Big)\Big].
\end{aligned}
\end{equation}  
Interestingly, though forms of both pairs of generators $X_4, X_5$ given in Eq. (15) and $G_4, G_5$ in (16) are different, even then they provide the same first integrals. Moreover, the Lie algebra of the operators $G_4$ and $G_5$ is Abelian, i.e., $[G_4, G_5]=0$ and the elimination of $u'$ from the Eq. (17) yields an analytical solution for the equation of under-damped oscillator as
\begin{equation}
u(t)=\frac{1}{\sqrt{4km-c^2}}\exp\Big(-\frac{ct}{2m}\Big)\Big[C_1\sin\Big(\frac{\sqrt{4km-c^2}t}{2m}\Big)-C_2\cos\Big(\frac{\sqrt{4km-c^2}t}{2m}\Big)\Big].
\end{equation}
\\ The commutation relations in view of operators (14) are displayed in TABLE II
\\
\\
\begin{table}[h!]
	\centering
	\begin{tabular}{c| c c c c c} 
		&$\textbf{X}_{1}$&$\textbf{X}_{2}$ & $\textbf{X}_{3}$ & $\textbf{X}_{4}$ & $\textbf{X}_{5}$ \\ 
		\hline
		\hline
		$\textbf{X}_{1}$ & $0$ & $\frac{\sqrt{4km-c^2}}{m}\textbf{X}_{3}$ & $-\frac{\sqrt{4km-c^2}}{m}\textbf{X}_{2}$ & $-\frac{\sqrt{4km-c^2}}{2m}\textbf{X}_{5}$ &$\frac{\sqrt{4km-c^2}}{2m}\textbf{X}_{4}$ \\
		
		$\textbf{X}_{2}$ &$-\frac{\sqrt{4km-c^2}}{m}\textbf{X}_{3}$&$0$& $-\frac{\sqrt{4km-c^2}}{m}\textbf{X}_{1}$ & $-\frac{\sqrt{4km-c^2}}{2m}\textbf{X}_{4}$ &$\frac{\sqrt{4km-c^2}}{2m}\textbf{X}_{5}$ \\
		
		$\textbf{X}_{3}$ &$\frac{\sqrt{4km-c^2}}{m}\textbf{X}_{2}$ & $\frac{\sqrt{4km-c^2}}{m}\textbf{X}_{1}$ & $0$ &$\frac{\sqrt{4km-c^2}}{2m}\textbf{X}_{5}$ & $\frac{\sqrt{4km-c^2}}{2m}\textbf{X}_{4}$\\
		
		$\textbf{X}_{4}$ & $\frac{\sqrt{4km-c^2}}{2m}\textbf{X}_{5}$ & $\frac{\sqrt{4km-c^2}}{2m}\textbf{X}_{4}$ & $-\frac{\sqrt{4km-c^2}}{2m}\textbf{X}_{5}$ &$0$ &$0$ \\
		
		$\textbf{X}_{5}$ & $-\frac{\sqrt{4km-c^2}}{2m}\textbf{X}_{4}$ & $-\frac{\sqrt{4km-c^2}}{2m}\textbf{X}_{5}$ & $-\frac{\sqrt{4km-c^2}}{2m}\textbf{X}_{4}$ &$0$ & $0$\\ [1ex]
	\end{tabular}
	\caption{Commutators Table}
	\label{table:2}
\end{table}\\
\textbf{Case 3. Critical-damped oscillator:}\\ If the restoring force and damping force are comparable in effect then the dynamical system behaves as critical damping oscillator. In this situation, we have $4km-c^2=0$ and the Eq. (5) provides the following 5-Noether symmetries
\begin{equation}
\begin{aligned}
\textbf{X}_{1}&=\exp\left(-\sqrt{\frac{k}{m}}t\right)\frac{\partial}{\partial u},\\
\textbf{X}_{2}&=t\exp\left(-\sqrt{\frac{k}{m}}t\right)\frac{\partial}{\partial u},\\
\textbf{X}_{3}&=\frac{\partial}{\partial t}-\sqrt{\frac{k}{m}}u\frac{\partial}{\partial u},\\
\textbf{X}_{4}&=t\frac{\partial}{\partial t}-\frac{1}{2}\Big(2t\sqrt{\frac{k}{m}}-1\Big)u\frac{\partial}{\partial u},\\
\textbf{X}_{5}&=\frac{1}{2}t^2\frac{\partial}{\partial t}-\frac{1}{2}\left(t^2\sqrt{\frac{k}{m}}-t\right)u\frac{\partial}{\partial u}.
\end{aligned}
\end{equation} 
We mention here that the forms of Noether symmetry generators in the critical damped harmonic case are entirely different from what we have seen for over-damped and under-damped cases.\\ The Lie algebra of these operators is shown below
\begin{table}[h!]
	\centering
	\begin{tabular}{c| c c c c c} 
		&$\textbf{X}_{1}$&$\textbf{X}_{2}$ & $\textbf{X}_{3}$ & $\textbf{X}_{4}$ & $\textbf{X}_{5}$ \\ 
		\hline
		\hline
		$\textbf{X}_{1}$ & $0$ & $0$ & $0$ & $\frac{1}{2}\textbf{X}_{1}$ &$\frac{1}{2}\textbf{X}_{2}$ \\
		$\textbf{X}_{2}$ &$0$&$0$& $-\textbf{X}_{1}$ & $-\frac{1}{2}\textbf{X}_{2}$ &$0$ \\
		$\textbf{X}_{3}$ &$0$ & $\textbf{X}_{1}$ & $0$ &$\textbf{X}_{3}$ & $\textbf{X}_{4}$\\
		$\textbf{X}_{4}$ & $-\frac{1}{2}\textbf{X}_{1}$ & $\frac{1}{2}\textbf{X}_{2}$ & $-\textbf{X}_{3}$ &$0$ &$\textbf{X}_{5}$ \\
		$\textbf{X}_{5}$ & $-\frac{1}{2}\textbf{X}_{2}$ & $0$ & $-\textbf{X}_{4}$ &$-\textbf{X}_{5}$ & $0$\\ [1ex]
	\end{tabular}
	\caption{Commutators Table}
	\label{table:3}
\end{table}\\
Now considering these Noether symmetries and employing the classical Noether's theorem provide respectively, the five first integrals 
\begin{equation}
\begin{aligned}
I_{1}&=\exp\Big(\sqrt{\frac{k}{m}}t\Big)\Big(\sqrt{\frac{k}{m}}u+u^{\prime}\Big),\\
I_{2}&=\exp\Big(\sqrt{\frac{k}{m}}t\Big)\Big(\sqrt{\frac{k}{m}}tu+tu^{\prime}-u\Big),\\
I_{3}&=\Big(\exp\Big(\sqrt{\frac{k}{m}}t\Big)\Big)^{2}\Big(\frac{ku^{2}}{2m}+\sqrt{\frac{k}{m}} uu^{\prime}+\frac{1}{2}u^{\prime 2}\Big),\\
I_{4}&=\Big(\exp\Big(\sqrt{\frac{k}{m}}t\Big)\Big)^{2}\Big(\frac{t}{2}\Big(\sqrt{\frac{k}{m}}u+u^{\prime}\Big)^{2}-\Big(\sqrt{\frac{k}{m}}tu-\frac{u}{2}+tu^{\prime}\Big)\Big(\sqrt{\frac{k}{m}}u+u^{\prime}\Big)\Big),\\
I_{5}&=\frac{u^{2}}{4}\exp\Big(2t\sqrt{\frac{k}{m}}\Big)+\frac{t}{4m}(2mu(tu^{\prime}-u)\sqrt{\frac{k}{m}}+tmu^{\prime 2}+tku^{2}-2muu^{\prime})\Big(\exp\Big(\sqrt{\frac{k}{m}}t\Big)\Big)^{2}.
\end{aligned}
\end{equation} 
We stress here that the forms of Noether symmetry generators $X_1, X_2, X_3, X_4$ given in Eq. (20) are similar to that found in \cite{Cer} and the first integrals as well. However, the form of operator $X_5$ is different and the resulting first integral is also new and fulfill the condition to becomes first integral of the critical damped harmonic oscillator. Furthermore, from the invariant quantities $I_1, I_2$ given in Eq. (20), we obtain following solution for the critical damped harmonic oscillator
\begin{equation}
u(x)=e^{-\sqrt{k/m}t}\Big[I_1 t-I_2\Big].
\end{equation}
Interestingly, we point out here that the insertion of the following Lagragian in Euler-Lagrange equation (3) also yields the same damped harmonic oscillator equation (8)
\begin{equation}
L=\frac{e^{{\frac{c}{m}}t}}{4m^2}\Big[2m^2u'^2+2cmuu'+(c^2-2km)u^2\Big].
\end{equation} 
Remarkably, if we employ this new form of Lagragian for the DHO in Noether symmetry condition and classical Noether's theorem, we obtain same set of Noether symmetries and constants of motion as we have determined for over-, under- and critical-damped oscillators. So this Lgargian can also be considered as the standard Lagragian. Though the new form of Lagrangian is completely different yet it yields same results as we derived using the Bateman Lagrangian. So this new form of optimal Lagragian function given in (22) may conceal something interesting for the physicists. We conclude here that this new form of Lagrangian may lead to entirely different quantum mechanical properties related to damped harmonic oscillator. Moreover, some interesting results related to DHO employing Eq. (22) will be discussed in future work.  
\section{Conclusion}
In this paper we have classified the harmonic oscillator problem with respect to Noether symmetries by constructing related conserved quantities. We have discussed three cases of damped harmonic oscillator namely, over-damped, under-damped and critical-damped. We observe that when we impose over-damped, under-damped and critical-damped conditions, the forms of few generators given in Eq. (10) get changed and provide new first integrals compared to that mentioned in Eq (11). Being an old and classical problem, several authors have attempted to study the group theoretic properties of variational DHO equation. In the current study we have also investigated the DHO equation of motion from group theoretic view point. In this paper we have determined some new results along with old one which have already been reported in the literature. For instance, in the over-damped harmonic oscillator case, the forms of all generators except $X_1$ are entirely different to what mentioned in \cite{Cer} and consequently the corresponding conserved quantities are also different. Similarly, for the under-damped harmonic oscillator case the real and imaginary parts of operator $X_4$ provide two Noether symmetry generators. These resulting operators form an Abelian algebra and yield two functionally independent first integrals. Furthermore, these two integrals suffice to provide the exact solution of under-damped oscillator. In the critical case the four Noether symmetry generators are similar to that mentioned in \cite{Cer}, while the operator $X_5$ is new and distinct from the list provided in \cite{Cer} and the resulting first integral also fulfills the condition to become the first integral of critical damped harmonic oscillator. In all three cases of damped harmonic oscillator the algebra of resulting Noether symmetry generators is closed. \\
We have also presented a new form of a Lagrangian function for the DHO system. We observe that for DHO on using new form of Lagrangian provides same set of Noether symmetries and associated conservation Laws as previously discussed for over-, under- and critical-damped oscillators. So this Lagrangian can also be considered as the standard Lagrangian. Though the new form of Lagrangian is completely different to that Bateman Lagrangian but it yields same results as we derive employing Bateman Lagrangian. So this new form of optimal Lagrangian function given in (22) may be usefully applied to explore some mechanical properties related to DHO.

\end{document}